\documentstyle[12pt,aaspp4]{article}
 
\input{epsf}

\journalid{}{}
\articleid{}{}
\begin{document}
\def\lax    {\ifmmode{_<\atop^{\sim}}\else{${_<\atop^{\sim}}$}\fi}
\def\gax    {\ifmmode{_>\atop^{\sim}}\else{${_>\atop^{\sim}}$}\fi}
\def\gtorder{\mathrel{\raise.3ex\hbox{$>$}\mkern-14mu
             \lower0.6ex\hbox{$\sim$}}}
\def\ltorder{\mathrel{\raise.3ex\hbox{$<$}\mkern-14mu
             \lower0.6ex\hbox{$\sim$}}}
 
\long\def\***#1{{\sc #1}}
 
\title{Detection of a 5-Hz QPO from X-ray Nova GRS~1739-278}

\author{K. N. Borozdin\altaffilmark{1,2}, S. P. Trudolyubov\altaffilmark{2,1}}

\altaffiltext{1}{NIS-2, Los Alamos National Laboratory,
Los Alamos, NM 87545}

\altaffiltext{2}{Space Research Institute, Russian Academy of Sciences, 
ul. Profsoyuznaya 84/32, Moscow, 117810 Russia}

\begin{abstract}

The X-ray nova GRS~1739-278 flared up near the Galactic center
in the spring of 1996.  Here we report on the discovery of a $\sim$5-Hz 
quasi-periodic oscillations (QPO) in RXTE/PCA observations of GRS~1739-278.  
The QPO were only present when the source was in its very high
state, and disappeared later, when it made a transition down into
the high state. We present the energy spectra of this black hole candidate 
measured in both high and very high states, and discuss the similarities 
between this system and other X-ray transients.

\end{abstract} 


\section{INTRODUCTION}

X-ray novae are outbursts of X-ray emission, often from black hole candidates,
that exceed their quiescent flux levels by many orders of magnitude. 
About 50 X-ray novae have been detected in the last 35 years, each lasting 
typically several hundred days before their return back to quiescence
(\cite{chen97}).  Perhaps all of them are recurrent, but the recurrence period 
is unknown for most. Probably, just small percentage of such systems
in the Galaxy have been discovered so far.

During outburst, X-ray novae
are typically found in one of several distinguishable 
spectral states(e.g. \cite{sun94,tl95,tsh96}). 
In the {\it high} state, the spectrum is composed of  
a bright thermal component and an extended steep power-law
with photon index around $\sim$2.5.
In the {\it low} state, the spectrum is a hard power-law 
with a photon index  of $\sim$1.5 and an exponential high energy cutoff. 
A third state, the so-called {\it very high} state has also been recognized.
It has a two component spectrum similar to the {\it high} state, 
but with somewhat stronger power-law component and with much more prominent 
fast variability (\cite{miya91,tak97}). Many X-ray novae
demonstrate state transitions during the outburst that are
associated with X-ray flux changes.
It is widely believed that both the state transitions and 
the flux variations are regulated by variations in the accretion rate. 
Similar states and state transitions were also observed in 
the persistent black hole systems, namely, Cyg X-1 and GX 339-4, but the
dynamics of these systems is much slower, so they are more often 
observed in one of these states for extended period of time
with occasional unpredictable transitions (\cite{sun79,mak86,dove98,Trud98}).

A new hard X-ray source GRS 1739-278 was discovered near the Galactic
Center on March 18, 1996 by the SIGMA/Granat gamma-ray telescope 
(Paul et al. 1996). Initial SIGMA localization was refined 
by the TTM/Kvant instrument (Borozdin et al. 1996). VLA radio
observations revealed a radio source within the TTM error region (Durouchoux
et al. 1996). Mirabel et al. (1996) measured the optical/infrared flux from
this object.       

In 1996, the source was observed in the X-ray band by ROSAT (Greiner et al.
1997), Granat (Vargas et al. 1997), RXTE (Takeshima et al. 1996), and the
Kvant module of the Mir Space Station.  Borozdin et al. (1998) presented
the spectral analysis of Mir-Kvant and RXTE data, and classified the
GRS 1739-278 as a soft X-ray nova and a black-hole candidate. In this 
letter, we report on the discovery of a 5-Hz QPO in the power density
spectrum, which supports the classification of this system as a binary
of black hole with low-mass companion.

\section{OBSERVATIONS AND DATA REDUCTION}

The RXTE satellite observed X-ray nova GRS 1739-278 on March 31, 1996
and nine more times from May 10 through May 29 of that year, each with an
exposure of several kiloseconds. The total exposure was about 24 ksec.   

The RXTE satellite has two co-aligned spectrometers with $\sim$1 degree 
field of view each: a set of five xenon proportional counters, PCA, 
with maximum sensitivity in 4-20 keV energy range; and a scintillation
spectrometer, HEXTE, which consists of eight NaI(Tl)/CsI detectors 
that are sensitive to 15-250 keV photons. 
HEXTE detectors are combined into two independent 
clusters of four detectors each, which alternate between measuring
the source and the X-ray background.

We used PCA {\em Binned} and {\em Single Binned} 
mode data in our timing analysis. We generated power density spectra 
(PDS) in the 0.001--256 Hz frequency range (2--13 keV energy band).
For lower frequencies (below 0.3 Hz) a single Fourier transform 
on the data binned in $0.125$ s time intervals was performed.
For higher frequencies we summed together the results of Fourier
transforms made over short stretches of data with $0.002$ s time bins.
The resulting spectra were logarithmically rebinned when necessary 
to reduce scatter at high frequencies and normalized to square root 
of $rms$ fractional variability.  The standard technique of subtracting
the white--noise level due to the Poissonian statistics, corrected 
for the dead--time effects, was employed (\cite{vgch94,zhang95}).
For spectral analysis we used FTOOLS v.4.2 and the PCA response matrix v.3.3
(see Jahoda et al. 1997 for computations of the matrix and 
Stark et al. 1997 for simulations of the background).

\section{LIGHT CURVE OF THE SOURCE}

\begin{figure}
\epsfxsize=17cm
\epsffile{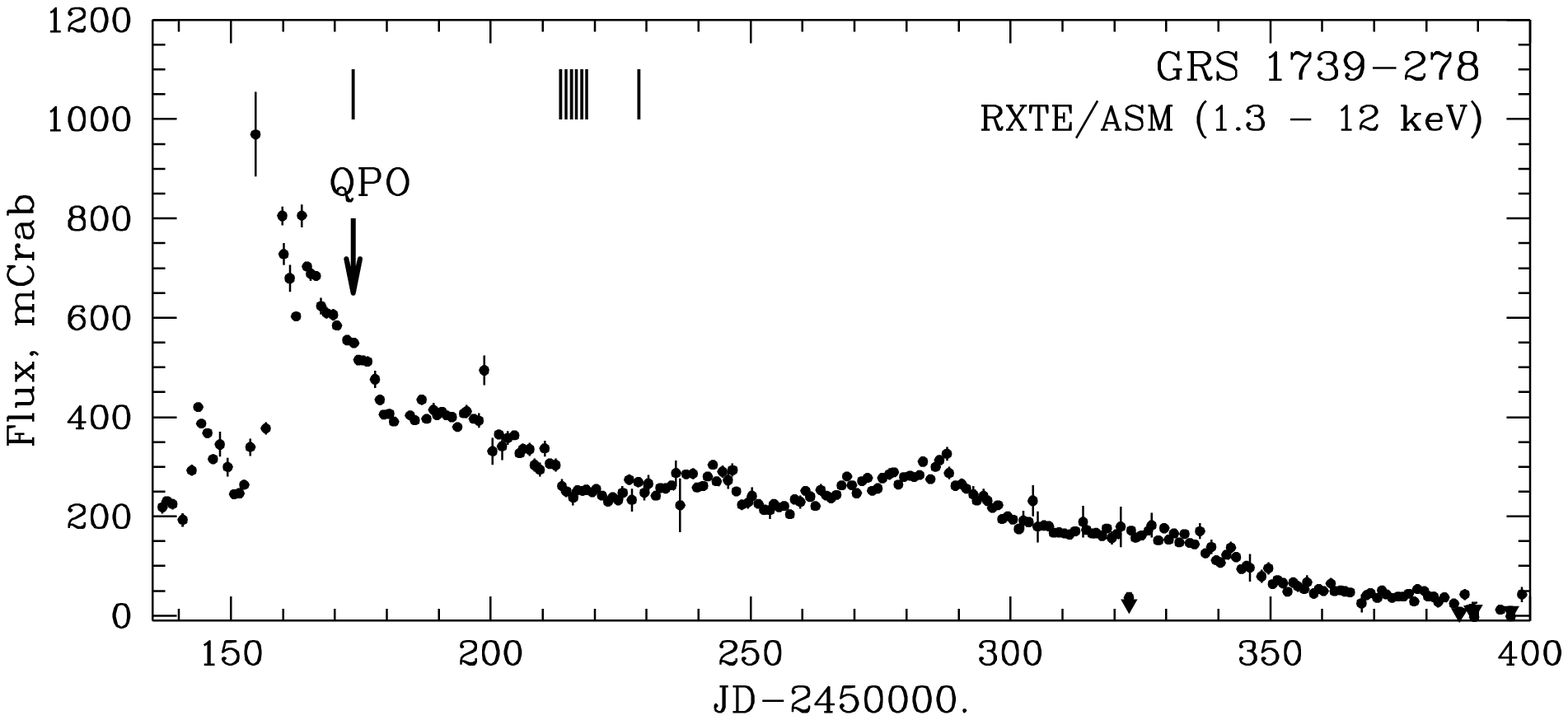} 
\caption{The light curve of GRS 1739-278, according to 
ASM/RXTE data (1.3-12 keV).  Tick marks indicate the
times of pointed RXTE observations. The arrow denotes
the time of 5-Hz QPO detection.}
\label{lc}
\end{figure}

Fig.\ \ref{lc} shows GRS~1739--278 light curve as measured 
with RXTE all-sky monitor (ASM) during the 1996 outburst. 
Overall shape of the outburst may be characterized as 
fast-rise-exponential-decay (FRED - \cite{chen97}). However,
the rise of this outburst was not particularly fast, and decay
was interrupted by multiple secondary maxima. Secondary maxima 
were observed in FRED-type light curves of many X-ray novae
(\cite{sun94,tsh96}).

The RXTE pointed instruments observed GRS~1739--278 during the decay 
of the outburst. The first observation (March 31) was made during 
the initial decay after
the main maximum.  X-ray flux from the source was about 600 mCrab in
PCA band (2-30 keV). A series of observations were also performed in May
after the secondary maximum. By that time the source had faded 
to 250-300 mCrab. During all of the May observations 
GRS~1739--278 displayed very similar energy 
spectra and featureless PDS shapes.

\section{5-HZ QPO IN POWER DENSITY SPECTRUM}

\begin{figure}
\epsfxsize=19cm
\epsffile{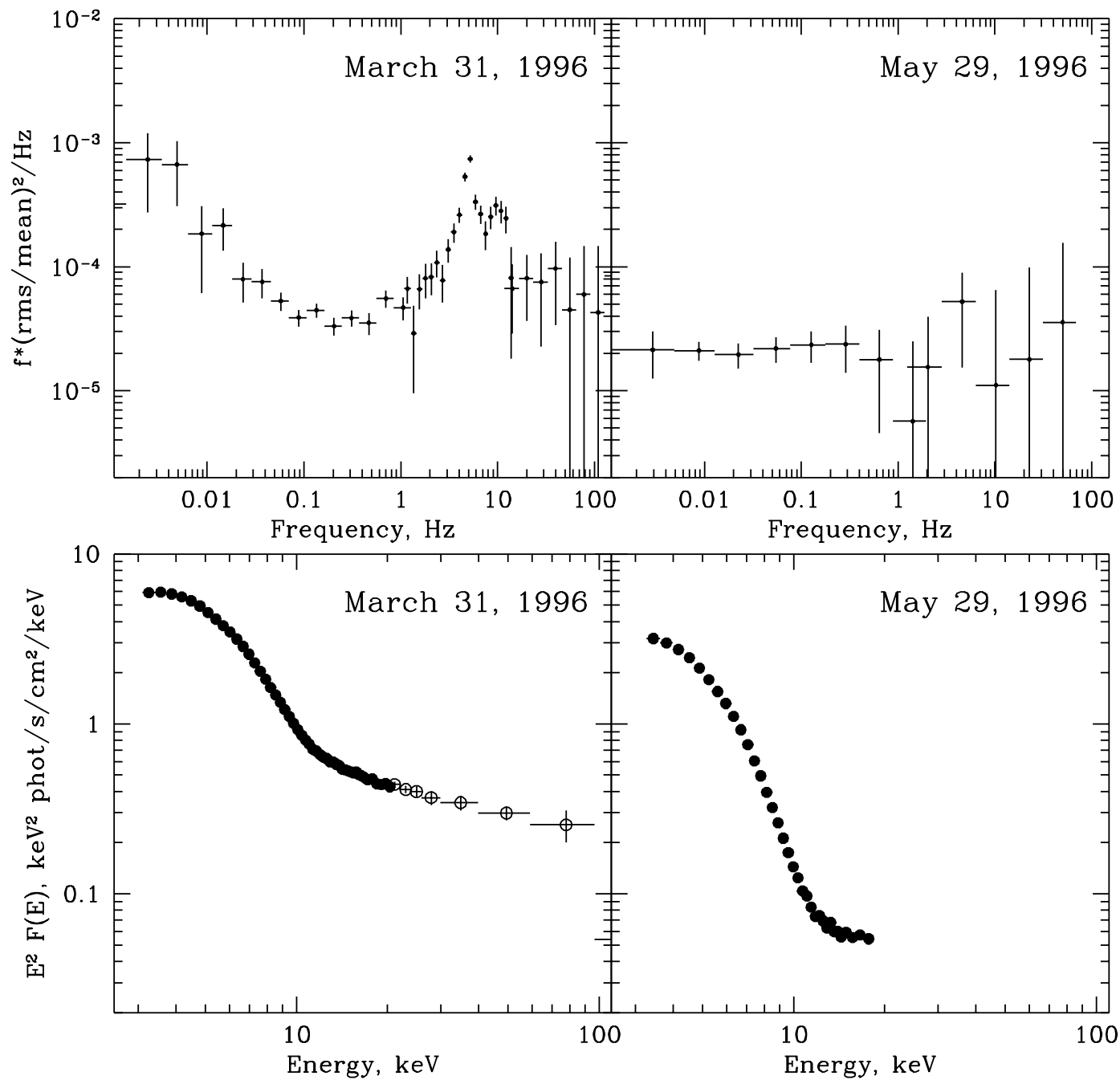} 
\caption{The results of PCA/RXTE observations of GRS~1739-278.
{\it Upper row}: The power density spectra (PDS) constructed
for the 2-13 keV band. {\it Lower row}: energy spectra.  
{\it Left column} is for the observation of March 31, 1996 (6.3 ksec), 
when 5-Hz QPO was observed.
{\it Right column} presents PDS and energy spectrum typical
for the series of observations in May 1996 (2.5 ksec exposure).
}
\label{pds_spectra}
\end{figure}

\begin{figure}
\epsfxsize=17cm
\epsffile{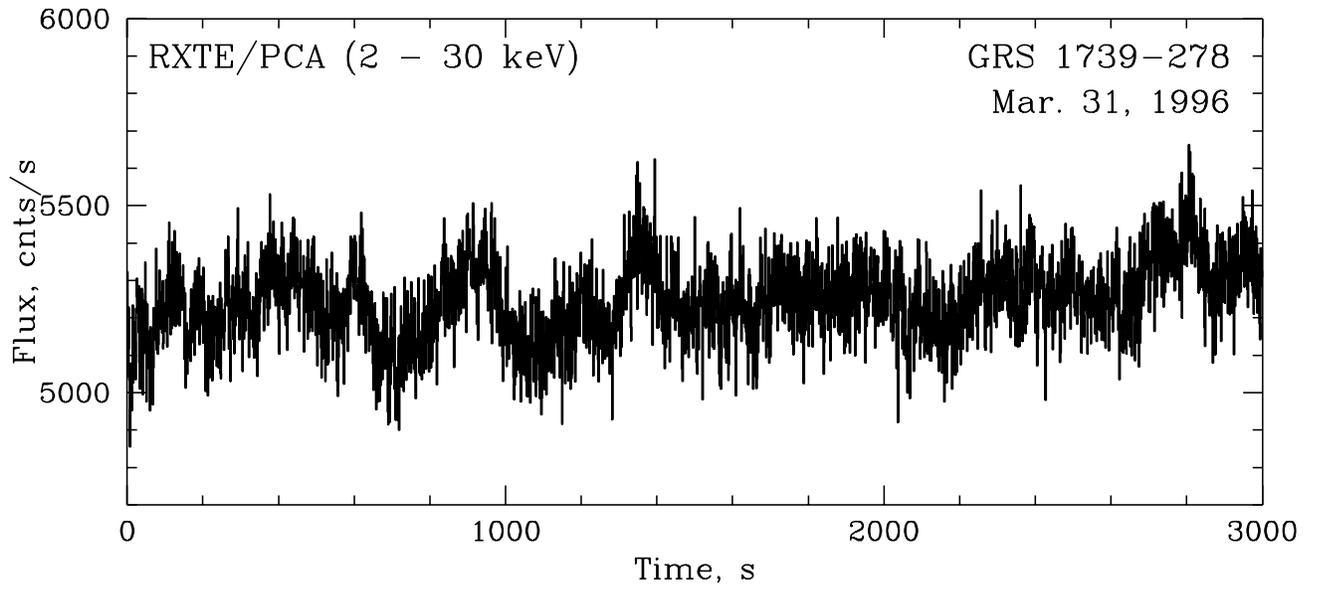} 
\caption{
The PCA light curve of GRS 1739-278 for March 31, 1996 observation
(2-30 keV). X-ray flux variations at time scales of tens of seconds
are clearly seen.
}
\label{dip_lc}      
\end{figure}

Construction of the PDS for the first PCA/RXTE observation (March 31, 1996 - 
Fig.\ \ref{pds_spectra}) revealed the presence of a QPO feature
with central frequency near 5 Hz (see Table 1 for the QPO parameters).
The QPO was seen clearly in 2-13 keV band, but was not significant
at higher energies, where the number of photons was small and hence 
the errors were larger. Also present in the PDS was a band-limited noise 
component, and significant variability at low frequencies.  
The variability of the source during the March 31 observation 
is shown in Fig.\ \ref{dip_lc}.
In contrast, much weaker fast variability was detected in 
the RXTE observations of the same source in May 1996
(Fig. \ref{pds_spectra}). 

\begin{figure}
\epsfxsize=16cm
\epsffile{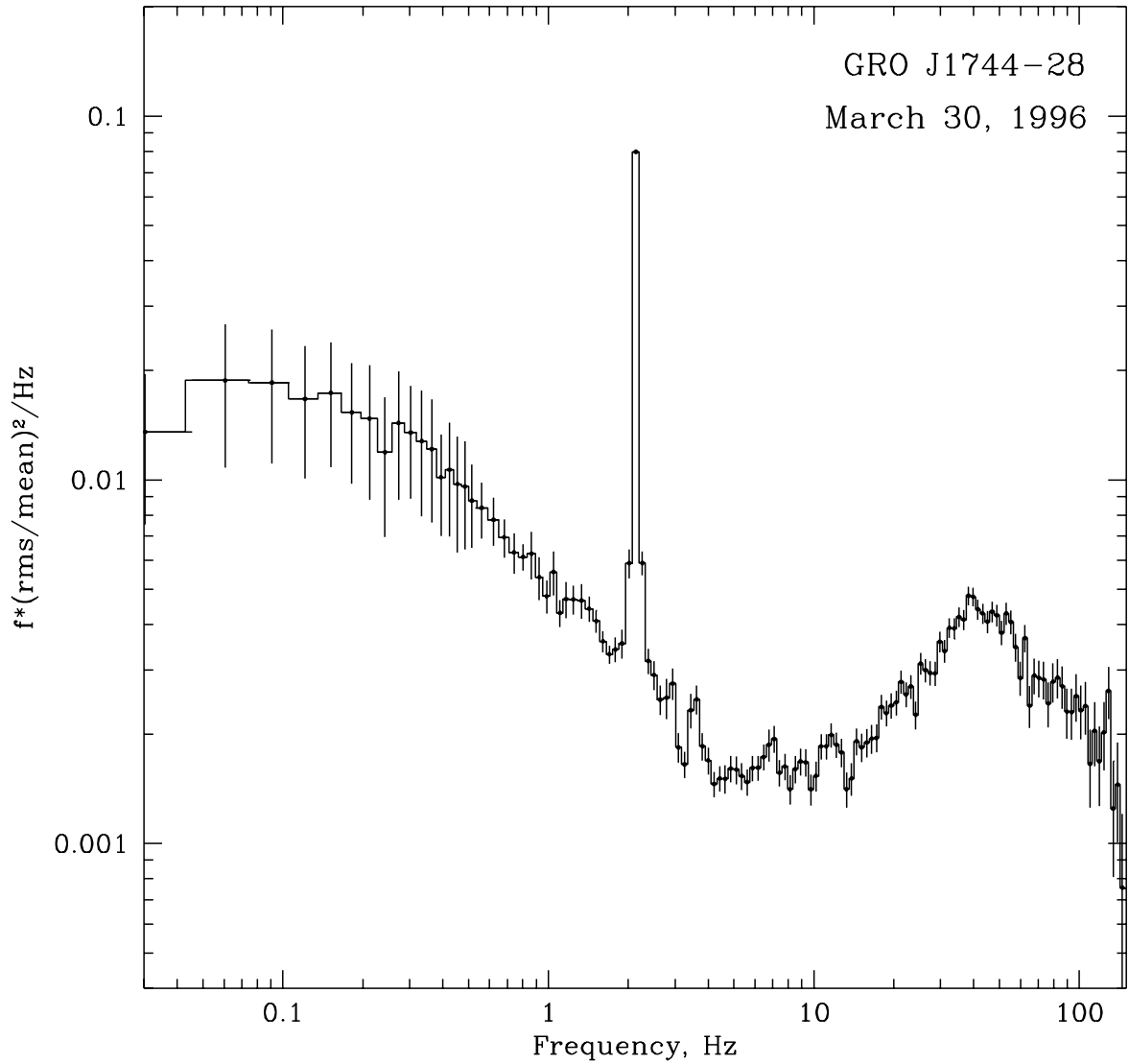} 
\caption{Power density spectrum of GRO~J1744--28 for
the PCA observation of March 30, 1996.
Dominating peak at $\sim$2 Hz corresponds to the pulsar
period.  The absence of a noticeable peak
at this frequency in the PDS of GRS~1739--278 (Fig.\ \ref{pds_spectra})
demonstrates that a contamination from GRO~J1744--28, if any, was
not significant factor.
}
\label{pds_1744}
\end{figure}

The pointing direction for the observations of GRS~1739--278
was offset by about 11 arcmin in order to reduce count rate
from the pulsar-burster GRO~J1744--28 (\cite{tak96}). However,
we were still concerned about the possible contamination of 
our power density spectra by this bright nearby source.
So we analyzed the data from its observation of March 30, 1996, just
one day before the first observation of GRS~1739--278 with RXTE took 
place.  The flux from GRO~J1744--28 in the PCA band (2-30 keV) was 925 mCrab, 
while for GRS~1739--278 (next day) it was only 604 mCrab.  
We built a PDS for GRO~J1744--28 to compare it with PDS of GRS~1739--278.
The result is presented in Fig.\ \ref{pds_1744}.  
A prominent peak at $\sim$2 Hz corresponding to the pulsar period 
dominates the PDS of GRO~J1744--28.  But there is no indication
of a $\sim$2 Hz peak in the PDS we derived for GRS~1739--278.  
So we conclude that contamination
of the GRS~1739--278 observations by GRO~J1744--28, if any, was not 
a significant factor, and that the detected $\sim$5 Hz QPO 
do belong to GRS~1739--278.

\begin{table}
\caption{Fit parameters for PDS components of GRS~1739-278\label{pds_fit}}
\begin{tabular}{cccc}
\hline
\multicolumn{4}{c}{\it Band-Limited Noise}\\
\hline
 & power-law index & break frequency, Hz & total $rms$ ($0.03 - 100$ Hz)$^{a}$\\
\hline
3/31/96 & $-1.22\pm0.22$ & $7.3\pm3.0$ & $1.9\pm0.1\%$ \\
5/29/96 & & & $0.4\pm0.2\%$ \\
\hline
\multicolumn{4}{c}{\it QPO and its harmonic}\\
\hline
 & centroid frequency, Hz & width, Hz & $rms$ amplitude\\
\hline
3/31/96 & $5.01\pm0.04$ & $1.17\pm0.18$ & $1.62\pm0.13 \%$ \\
        & $9.80\pm0.35$ & $3.18\pm1.07$ & $1.17\pm0.22 \%$ \\
\hline
\end{tabular}
\par
$^{a}$ -- integrated {\em total} $rms$ of fractional variability 
in the $0.03 - 100$ Hz frequency range 
\end{table}

The energy spectra for all RXTE observations of GRS~1739--278 
(see examples in Fig.\ref{pds_spectra})
have the shape which is typical for X-ray novae (\cite{tsh96}). In
general, such spectra are well fitted by a two-component model 
composed of "multicolor" accretion disk component
(Makishima et al. 1986) in the soft part of the spectrum with a power-law
component at higher energies.  Detailed spectral analysis of GRS~1739--278
observations was presented by Borozdin et al. (1998, 1999).
During the observation on March 31, 1996, when the QPO was detected,
the power-law component in the energy spectrum was more prominent.
During the subsequent observations in May 1996 the QPO was not detected,
the power-law component waned, and no hard flux was detected by HEXTE.

Strong rapid variability in X-ray band and extended power-law component
in energy spectrum are both features of black hole binaries, when in
{\it very high} state (\cite{miya91,klis95}). 
We see that GRS~1739--278 was in this state
on March 31 1996, and made a transition down to a {\it high} state 
sometime before May 10.

\section{DISCUSSION}

QPO features in power density spectra have been identified
for a variety of black hole candidates including
GX~339--4 (\cite{miya91}), Nova Muscae (\cite{miya93,ebi94}),
XTE~J1748-288 (\cite{rev1748}), and 4U~1630--47 (\cite{trud1630}). 
Significant low-frequency variability has been observed in Galactic
microquasars GRS~1915+105 (\cite{Morgan97}) and GRO~J1655--40 
(\cite{Remillard99}), and also during the 1998 outburst of recurrent 
X-ray Nova 4U~1630--47 (\cite{trud1630}).
All of these objects were in their 
{\it very high} states during those observations.
In many cases a correlation between
the intensity of PDS noise components and relative 
strength of the power-law spectral component
was reported (e.g. \cite{miya93,cui99,rev1748,trud99}).  
GRS~1739--278 fits well into this picture
as another example of transient LMXB (low-mass X-ray binary)
and a black hole candidate.

An interesting energy spectrum was observed with TTM telescope on
Mir-Kvant module in March 1996 during the rise of the flux from the
GRS~1739--278 (\cite{bor98}). It is well described by a power law 
with absorption and does not
require the introduction of an additional soft blackbody component. At the
same time, the slope of the power-law component (2.3-2.7) is much steeper
than the typical value for the {\it low} state of 
black-hole candidates (1.5-2.0).
A similar spectrum was observed earlier by Ginga and Granat satellite 
from the X-ray Nova Muscae 1991 (\cite{greb91,gil91,ebi94})
and by TTM/Kvant and SIGMA/Granat from KS 1730-312 (\cite{bor95,trud96}).  
Later this type of spectrum 
was observed with RXTE from XTE~J1748-288 (\cite{rev1748}),
4U~1630-47 (\cite{trud1630}), and XTE~J1550-564 (\cite{sob99}).
These examples show that the power-law shape of the spectrum 
with a variable slope is typical of soft X-ray Novae during 
their flux rise and near the primary maximum. 
There is a clear tendency for the spectrum to steepen as 
the outburst progresses. However, the total X-ray luminosity is sometimes
even higher than in the {\it very high} state observed 
later during the same outburst (\cite{rev1748,trud1630}).
The observation of power law spectrum with variable slope
at the early phase of X-ray novae outbursts is important because
this is when an accretion disk around black hole is formed.
The generation of such spectrum should be a key
element of any sound dynamical model for an accreting black hole.

\section{CONCLUSION}

Using PCA/RXTE archival data we detected, for the first time, a QPO
feature in the PDS of X-ray nova GRS~1739-278. QPO harmonics near 10 Hz
and strong band-limited noise at low frequencies were also observed.
Both the PDS and the two-component energy spectrum for this observation
displayed the properties typical for the
{\it very high} state in black hole candidates.

In the later stages of the 1996 outburst, GRS~1739-278 transitioned into
a {\it high} state with much weaker rapid variability and soft energy
spectra that were dominated by thermal component with an effective temperature
$\sim$1 keV.  Observation of the two canonical states strongly supports
the identification of GRS~1739-278 as a black hole candidate based on
the similarity of its X-ray properties with other black hole
X-ray binaries.

We note, that at the beginning of the outburst GRS~1739-278 exhibited
a power-law energy spectrum with a variable slope and tendency 
to steepen with time (\cite{bor98}).
Similar spectra have been measured from several other
X-ray novae during the early stages of their outbursts.
These black hole candidates therefore seem to display 
a clear pattern in their spectral evolution.

\section{ACKNOWLEDGMENTS}

The RXTE public data were extracted from the HEASARC
electronic archive operated by the Goddard Space Flight Center (NASA).
We used also quick-look results provided by the ASM/RXTE team.
Authors are grateful to Dr.\ Thomas Vestrand for his valuable suggestions,
which helped us to improve the paper.

\end{document}